# Uranium polyhydrides at moderate pressures: prediction, synthesis, and expected superconductivity

**Authors:** I. A. Kruglov[1,2,*], A. G. Kvashnin[3,2], A. F. Goncharov[4,5,*], A. R. Oganov[3,2,1,*], S. Lobanov[5,6], N. Holtgrewe[5,7], Sh. Jiang[4], V. Prakapenka[7], E. Greenberg[7], A. V. Yanilkin[1,2].

**Affiliations:**

[1]Dukhov Research Institute of Automatics (VNIIA), Moscow 127055, Russian Federation

[2]Moscow Institute of Physics and Technology, Dolgoprudny, Moscow Region 141700, Russian Federation

[3]Skolkovo Institute of Science and Technology, 3 Nobel St., Moscow 143026, Russian Federation

[4]Key Laboratory of Materials Physics, Institute of Solid State Physics CAS, Hefei 230031, China

[5]Geophysical Laboratory, Carnegie Institution of Washington, Washington, DC 20015, USA

[6]V.S. Sobolev Institute of Geology and Mineralogy SB RAS, Novosibirsk 630090, Russian Federation

[7]Center for Advanced Radiations Sources, University of Chicago, Chicago, Illinois 60637, USA

**Abstract:** Hydrogen-rich hydrides attract great attention due to recent theoretical (*1*) and then experimental discovery of record high-temperature superconductivity in $H_3S$ ($T_c$ = 203 K at 155 GPa (*2*)). Here we search for stable uranium hydrides at pressures up to 500 GPa using *ab initio* evolutionary crystal structure prediction. Chemistry of the U-H system turned out to be extremely rich, with 14 new compounds, including hydrogen-rich $UH_5$, $UH_6$, $U_2H_{13}$, $UH_7$, $UH_8$, $U_2H_{17}$, and $UH_9$. Their crystal structures are based on either common f.c.c. or h.c.p. uranium sublattice and unusual $H_8$ cubic clusters. Our high-pressure experiments at 1-103 GPa confirm the predicted $UH_7$, $UH_8$, and three different phases of $UH_5$, raising confidence about predictions of the other phases. Many of the newly predicted phases are expected to be high-temperature superconductors. The highest-$T_c$ superconductor is $UH_7$ predicted to be thermodynamically stable at pressures above 22 GPa (with $T_c$ = 44-54 K) and this phase remains dynamically stable upon decompression to zero pressure (with $T_c$ = 57-66 K).

**Introduction**

Uranium hydride is a highly toxic compound spontaneously igniting in air (*3*) and reacting with water (*4*). It is used mainly for separation of hydrogen isotopes (*5*), but it can also be a component of explosives. For the first time it was synthesized by F. Driggs during heating of metallic uranium in hydrogen atmosphere, and it was initially assigned composition $UH_4$. Later the composition was determined as $UH_3$ (*6*): this phase is known as β-$UH_3$. When bulk uranium was heated in hydrogen atmosphere at lower temperature (*7*), a metastable α-$UH_3$ phase appeared and transformation to β-$UH_3$ occurred above 523 K. The crystal structure of α-$UH_3$ is of $Cr_3Si$-type (also known as A15 or β-W); U atoms have icosahedral coordination and form a b.c.c. sublattice. Many superconductors belong to the $Cr_3Si$ structure type (e.g. $Nb_3Sn$). The structure of β-$UH_3$ is more complex and based on a β-W-like uranium sublattice containing hydrogen atoms in distorted tetrahedral voids (*8*, *9*). $UH_3$ is the only known uranium hydride found to be stable at ambient conditions (though there is evidence of isolated molecules of UH, $UH_2$, $UH_4$, $UH_3$, $U_2H_2$, and $U_2H_4$ (*10*)). Hydrogen, being a molecular solid, has large atomic volume in the elemental form – volume reduction, favorable under pressure, can be achieved through compound formation, and one expects polyhydrides to form under pressure. Another motivation was that some uranium compounds were shown to have peculiar (and previously inconceivable) coexistence of superconductivity and ferromagnetism (*11*–*13*), and (non-magnetic) polyhydrides are prime candidates for high-temperature superconductivity.

**Results and discussion:**

In order to predict stable phases in the U-H system at pressures of 0, 5, 25, 50, 100, 200, 300, 400, and 500 GPa, we performed variable-composition evolutionary structure/compound searches using the USPEX algorithm (*14*–*16*). By definition, a thermodynamically stable phase has lower Gibbs free energy (or, at zero Kelvin, lower enthalpy) than any phase or phase assemblage of the same composition. Having predicted stable compounds and their structures at different pressures (Fig. 1), we built the composition-pressure phase diagram (see Fig. 2) which shows pressure ranges of stability for all the phases found. As shown in Fig. 2, our calculations correctly reproduce stability of both phases of $UH_3$ and predict 12 new stable phases corresponding to 14 new compounds (*Cmcm*-UH and $P6_3/mmc$-UH, *Ibam*-$U_2H_3$, *Pbcm*-$UH_2$, $C2/c$-$U_2H_5$, α-$UH_3$ and β-$UH_3$, *Immm*-$U_3H_{10}$, $P6_3mc$-$UH_5$, $P6_3/mmc$-$UH_6$, $P\bar{6}m2$-$U_2H_{13}$, $P6_3/mmc$-$UH_7$, $Fm\bar{3}m$-$UH_8$, $P\bar{4}m2$-$U_2H_{17}$, $P6_3/mmc$-$UH_9$ and metastable $F\bar{4}3m$-$UH_9$). Detailed information on crystal structures of the predicted phases can be found in Supplementary Table S1 and Fig. S1. Phonon calculations confirmed that none of the newly predicted phases have imaginary phonon frequencies in their predicted ranges of thermodynamic stability (see phonon dispersion curves and densities of states in Supplementary Materials). The only uranium hydride phase remaining stable at zero pressure is β-$UH_3$ which transforms into α-phase above 5.5 GPa (SM Fig. S2).

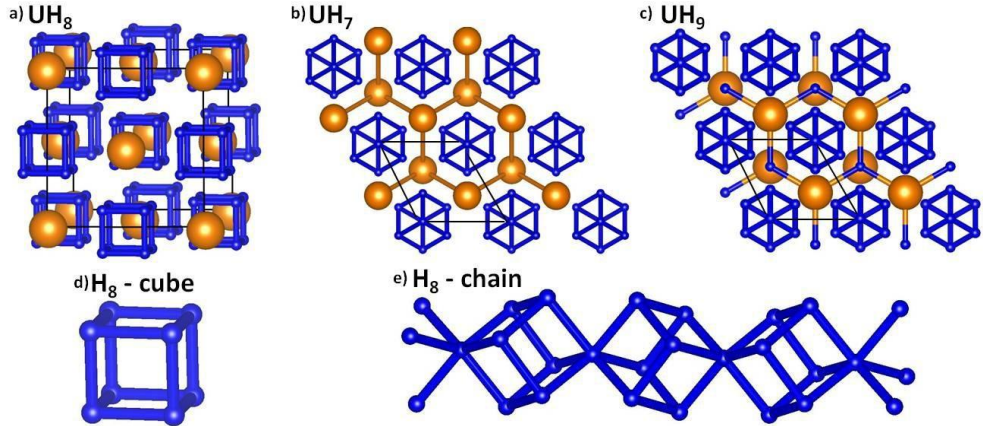

Figure 1: Crystal structures of the predicted a) $Fm\bar{3}m$-UH$_8$, b) $P6_3/mmc$-UH$_7$ and c) $P6_3/mmc$-UH$_9$ phases; d-e) basic hydrogen motifs. U atoms are shown by large orange balls and hydrogens by small blue balls.

Among the numerous predicted stable U-H phases, below we focus on hydrogen-rich (H/U > 3) polyhydrides as potential high-temperature superconductors. All these hydrogen-rich phases are metallic and their crystal structures feature very striking building blocks - H$_8$ cubes (Fig. 1d,e). Lattice dynamics calculations show a big gap between phonon contributions from uranium and hydrogen atoms, caused by their large mass difference: low-frequency modes mostly belong to uranium atoms, while high-frequency modes belong to hydrogen.

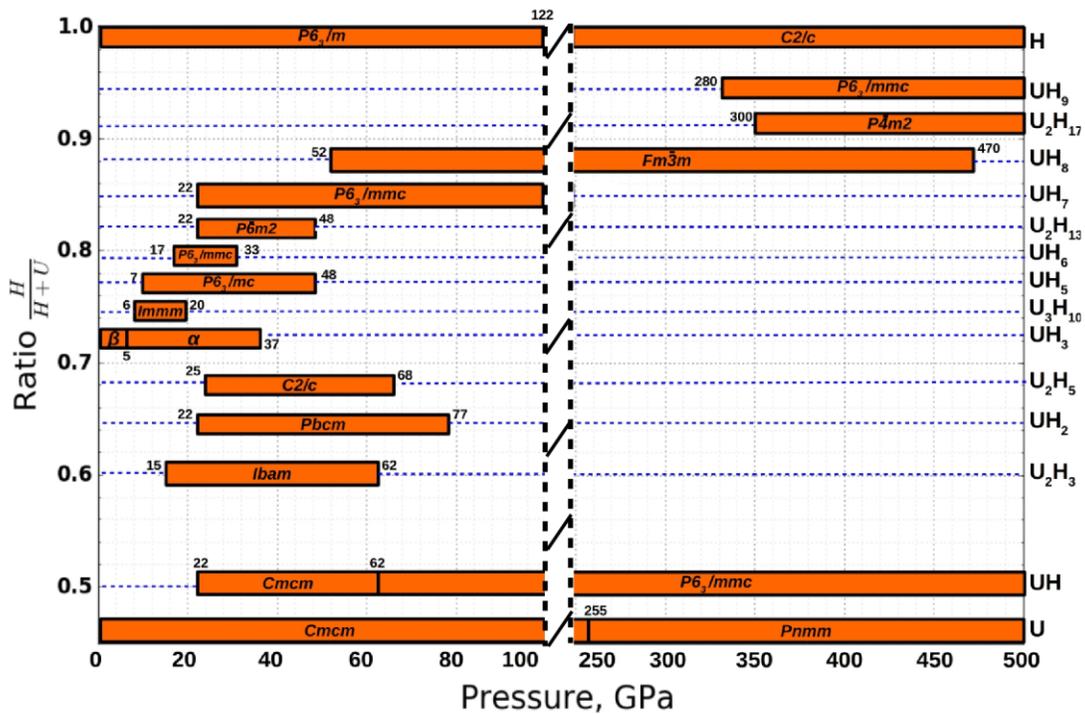

Figure 2: Pressure-composition phase diagram of the U-H system.

The first hydrogen-rich compound, $U_3H_{10}$, is predicted to become stable at 6 GPa. This structure is a derivative of α-$UH_3$ with one additional hydrogen atom and an enlarged lattice parameter along the $c$-direction. At pressures above 7 GPa, a family of $UH_{5-7}$ hydrides becomes stable with the same h.c.p. sublattice of U atoms. The $P6_3/mmc$-$UH_7$ structure (Fig. 1b) is a derivative of the anti-NiAs structure, where all the octahedral voids of U sublattice are filled by $H_8$-cubes (occupying the same positions as nickel atoms in the NiAs structure), forming infinite 1D-chains that run along the $c$-axis (Fig.1e). At 20 GPa, the shortest H-H distance is 1.56 Å and the U-H distances vary from 2.12 to 2.25 Å. $P\bar{6}m2$-$U_2H_{13}$ is a derivative of the $UH_7$ structure, with a doubled unit cell along the $c$-direction: U atoms still form an h.c.p. sublattice, but half of linkages between $H_8$ cubes are deleted. As a result, instead of infinite 1D-chains of corner-sharing pairs of $H_8$-cubes, we have isolated ("0D") corner-sharing pairs of incomplete $H_7$-cubes, and the composition changes from $UH_7 = U_2H_{14}$ to $U_2H_{14-1} = U_2H_{13}$. At 20 GPa, the H-H distance is 1.55 Å and the U-H distances vary from 2.09 to 2.13 Å. Derivatives of $U_2H_{13}$, the $UH_6$ and $UH_5$ structures form after deletion of corner-sharing H-atoms in the pair of $H_7$-cubes and another hydrogen atom from the chain, respectively. Evolution of the hydrogen chains in the octahedral voids is presented in Fig. 3. The only magnetic phase among $UH_{5-7}$ hydrides is $UH_5$ (Fig. S3). Strikingly, lattice dynamics calculations indicate that all these phases are dynamically stable even at zero pressure, i.e., they may exist as metastable materials at atmospheric pressure.

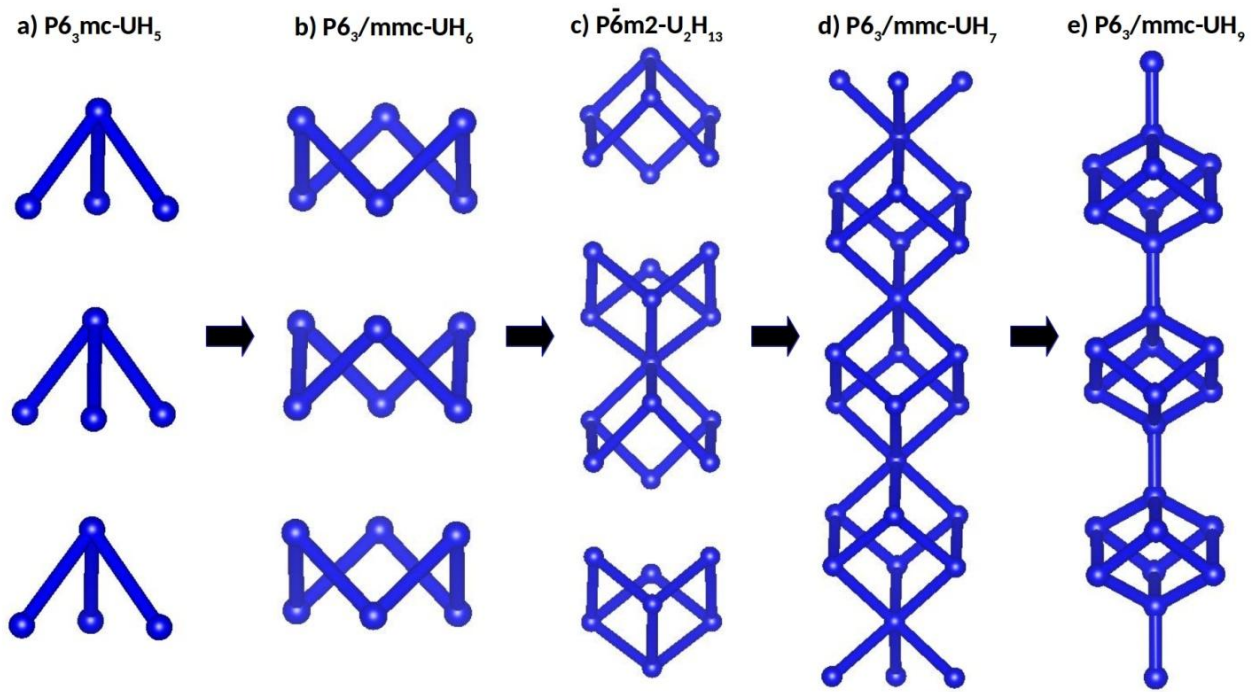

Figure 3: Evolution of hydrogen sublattice in h.c.p. $UH_{5-7}$ structures.

At a higher pressure of 52 GPa, a new stable hydrogen-rich phase is predicted – $UH_8$ (Fig. 1a). At pressures between 100 and 280 GPa, $UH_8$ is the only stable hydrogen-rich uranium hydride. Its structure is a derivative of the rocksalt structure, where U atoms occupy Na-sites and isolated $H_8$-cubes occupy Cl-sites. At 50 GPa, the H-H distances are 1.38 Å and the U-H distances are 2.13 Å. The coordination number of uranium atoms in this remarkable structure is equal to 24. At 280 GPa, another stable compound appears - $UH_9$ (Fig. 1c). It is structurally similar to $UH_7$: just like $UH_7$, it has an h.c.p. sublattice of uranium atoms and infinite 1D-chains of corner-sharing $H_8$-cubes running along the *c*-axis, but in addition, it has single H atoms located within close-packed uranium layers (Fig. 1c). At 300 GPa, the shortest H-H distances are 1.13 Å within $H_8$-cubes and 1.09 and 1.23 Å between the cubes. Besides the above-mentioned stable phase of $UH_9$, our calculations uncovered a low-enthalpy metastable (by just 18 meV/atom) polymorph with a space group $F\bar{4}3m$. This metastable structure (Fig. S1j) is based on the f.c.c. sublattice of uranium atoms, all the octahedral voids of which are occupied by isolated $H_8$ cubes-and half of the tetrahedral voids are occupied by single H atoms. This structure can be described as a half-Heusler alloy. Interestingly, USPEX found a stable $U_2H_{17}$ (Fig. S1i) whose structure is a derivative of the cubic $UH_9$, with half of single H atom positions vacant in the alternating layers – which lowers symmetry to a tetragonal $P\bar{4}m2$.

We have successfully synthesized the predicted uranium hydrides in a diamond anvil cell at various pressures (1-103 GPa) on loading and unloading cycles, applying moderate laser heating (up to 2000 K) on the loading cycles. Three experiments examined the reaction pathways of U and $H_2$; in one control experiment, U was loaded in Ar medium; in the other three, naturally oxidized uranium samples were studied in $H_2$ medium (these latter ones will not be considered here). The experiments with Ar medium revealed the presence of metallic U and a small amount (<5%) of f.c.c. $UO_2$ (e.g., Ref. (*17*)). In the experiments with $H_2$ medium, $UO_2$ was not detected. Shortly after $H_2$ loading to an initial pressure (0.1 to 2 GPa), the U sample swells and changes appearance, and XRD shows the formation of coexisting α-$UH_3$ and β-$UH_3$ (Fig. 4(a)). At the lowest pressures, the amount of α-$UH_3$ is approximately 30%. As the pressure increases, the share dwindles until this phase becomes undetectable above 5 GPa (or 7.5 GPa if not heated).

In contrast, β-$UH_3$ remains metastable up to 69 GPa (if not heated). The experimental unit cell volumes of these $UH_3$ phases are slightly larger (within 3%) compared to our theoretical predictions (Fig. 5). Above 5 GPa, a new phase starts to appear after laser heating. It becomes a dominant phase at 8 GPa (Fig. 4(b)) and remains detectable at up to 36 GPa. It is f.c.c.-based, and the hydrogen content can be evaluated using the measured and calculated unit cell volumes (Fig. 5). If we assume x=5, the experimental volumes are again slightly larger than the theoretically calculated ones. Other predicted $UH_5$ compounds (with very similar volumes) are h.c.p.-like and orthorhombic, and these will be discussed below. The structure of f.c.c. $UH_5$ is similar to the predicted $UH_8$, but instead of $H_8$ cubes it has alternating $H_4$ tetrahedra and single hydrogen atoms (Fig. S5b). At 31 GPa, another hydride starts appearing; it matches well the predicted h.c.p. $UH_7$ phase. This polyhydride can be observed at up to 103 GPa, which is the highest pressure in this study, and can be present as a single phase. However, above 45 GPa weak peaks of yet another f.c.c. structure appear (Fig. 4(c)), which remains stable up to 103 GPa. Volume of this phase is close to the theoretically found $F\bar{4}3m$-$UH_9$ structure (metastable above 280 GPa) (see Fig. 5), but its formation pressure is close to the $UH_8$ phase (52 GPa). Thus we believe that this experimental phase is an intermediate structure between f.c.c. UH8 and UH9 phases, and after in the text we will denote it as UH$_{8+\delta}$.

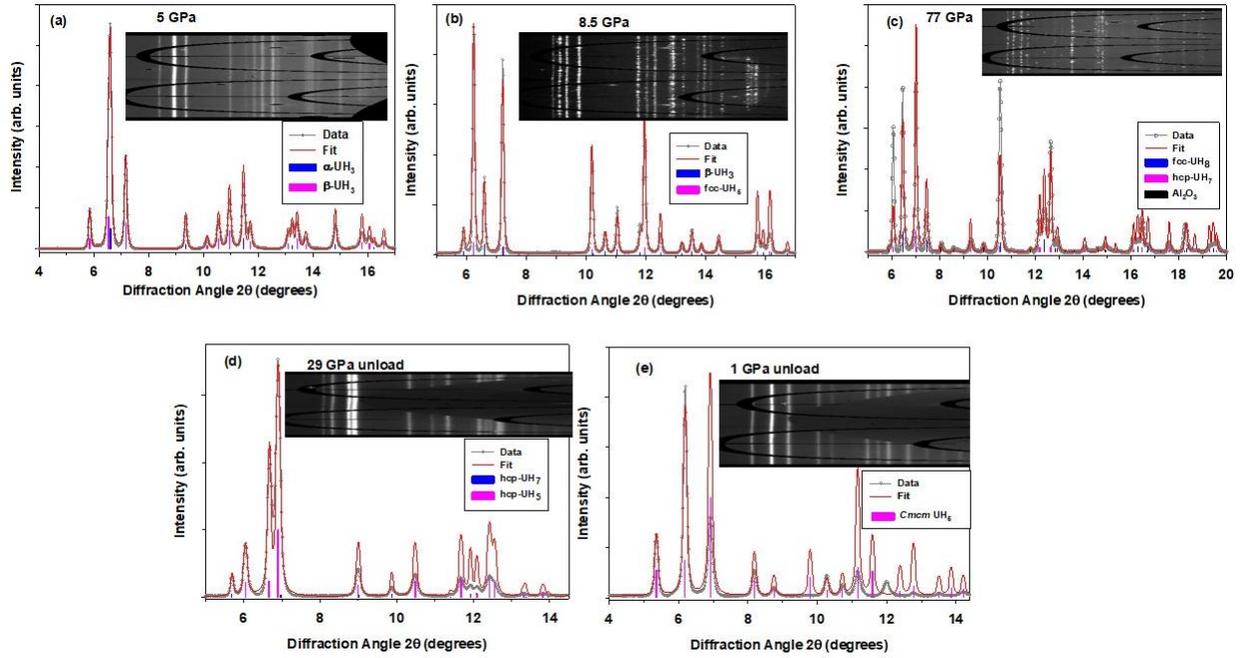

Figure 4: Experimental data on U-H compounds. XRD patterns of synthesized U-H phases: α-UH$_3$ and β-UH$_3$ (a); f.c.c. UH$_5$ (b), h.c.p. UH$_7$ and f.c.c. UH$_{8+\delta}$ (c), h.c.p. UH$_5$ (d), *Cmcm*-UH$_5$ (e).

We also looked at potential metastability of the phases synthesized at extreme conditions and also at possible appearance of some other metastable phases which can be realized because of kinetic reasons. No laser heating was applied in unloading cycles from 60, 45, and 38 GPa. A uniquely identified hexagonal UH$_7$ phase remains metastable down to at least 29 GPa. At this pressure we could have been able to identify a hexagonal UH$_5$ phase which was theoretically predicted (Fig. 4(d)). However, at lower pressure the diffraction patterns become very complex, possibly representing a mixture of several phases that could not be uniquely identified, and also remnants of unreacted β-UH$_3$. After unloading to nearly-ambient conditions (near 1 GPa), a single phase appeared to have been formed, which we were able to index in an orthorhombic lattice with a=3.438 Å, b=7.15 Å, and c=6.20 Å and the space group *Cmma* (Fig. 4(e)). The unit cell volume suggests the UH$_5$ composition (Fig. 5), although UH$_4$ could also be possible. The best match from among the theoretically predicted structures is *Cmcm*-UH$_5$ which is above the convex hull by 5 meV/atom at 5 GPa. This structure describes well the majority of the observed peaks, although there are some discrepancies too (Fig. 4(e)). The volumes-pressure data for the newly synthesized α-UH$_3$ and β-UH$_3$, orthorhombic UH$_5$, h.c.p. UH$_5$ and UH$_7$, and f.c.c. UH$_5$ and UH$_8$ are compared in Fig. 5 to the calculations and combinations of the experimental equations of state of h.c.p. H$_2$ and metallic U (*18*). The latter comparison shows that polyhydrides are stable when their volumes are lower than the sum of volumes of the elements. As one can see in Fig. 5, polyhydrides with higher H content require higher pressures for their stability - this trend emerges both from our experiments and structure predictions.

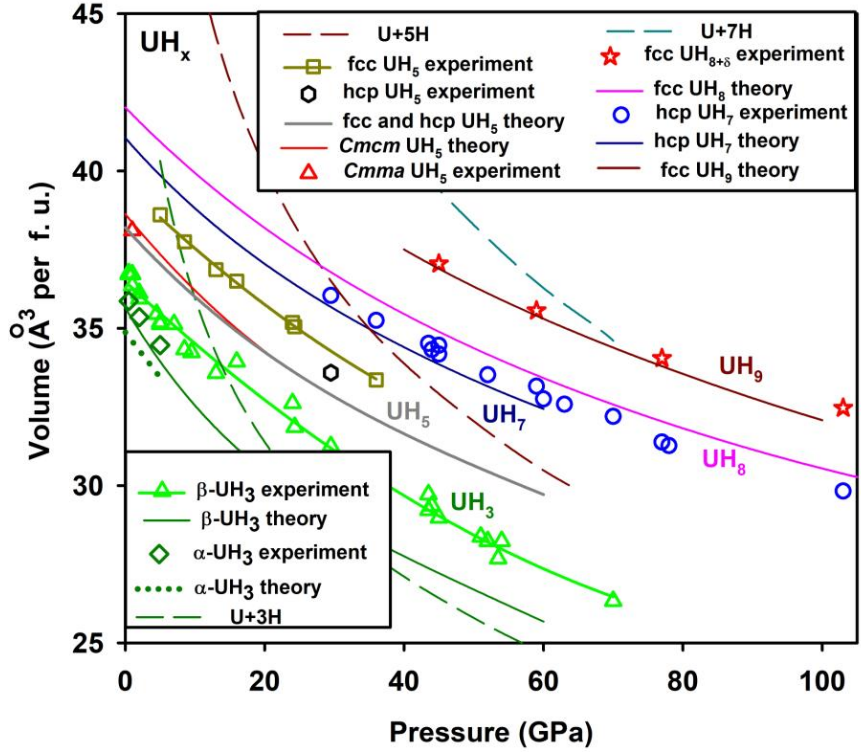

Figure 5. Volumes per formula unit as a function of pressure for the polyhydrides synthesized in this work in comparison to theoretical predictions. Also shown are the combined literature volumes of U metal and solid molecular hcp-$H_2$ in different proportions to illustrate the stability of polyhydrides at high pressures.

The calculated electronic band structures of predicted stable U-H phases are shown in Fig. S3. All these phases are metallic and feature numerous flat bands near the Fermi level. Only $UH_3$ and $UH_5$ phases are magnetic at low pressures (below 100 and 170 GPa, respectively). Their ferromagnetism can also be seen from the band structures shown in Fig. S3a-c, where contributions of U atoms to the spin-up and spin-down states are shown in red and blue, respectively.

While for UH3 the bands near the Fermi level come mainly from uranium orbitals, in uranium polyhydrides contributions of both uranium and hydrogen atoms are large. Geometric similarities (e.g., between $P6_3/mmc$-$UH_7$ and $P\bar{6}m2$-$U_2H_{13}$) result in similar densities of states and electronic properties, see Fig. S3 e-f. Likewise, stable $P6_3/mmc$ and metastable $F\bar{4}3m$ polymorphs of $UH_9$ (actually, these can even be described as polytypes) have similar electronic DOS due to geometric similarities of their crystal structures. As we will see below, similarities are observed also for electron-phonon coupling coefficients and superconducting $T_c$.

Table 1. Predicted superconducting properties of uranium hydrides. Two $T_c$ values calculated by solving the Eliashberg equation given for μ* equal to 0.1 and 0.15, respectively.

| Phase | Space group | P, GPa | $\omega_{\log}$, K | $\lambda$ | $T_c$, K |
|---|---|---|---|---|---|
| UH$_7$ | $P6_3/mmc$ | 20 | 873.8 | 0.83 | 54.1 / 43.7 |
| | | 0 | 764.9 | 0.95 | 65.8 / 56.7 |
| UH$_8$ | $Fm\bar{3}m$ | 50 | 873.7 | 0.73 | 33.3 / 23.4 |
| | | 0 | 450.3 | 1.13 | 55.2 / 46.2 |
| UH$_9$ | $P6_3/mmc$ | 300 | 933.4 | 0.67 | 31.2 / 19.9 |

In the Migdal-Eliashberg theory of superconductivity, the central quantity is the electron-phonon coupling (EPC) coefficient λ. The superconducting transition temperature ($T_c$) can be calculated using the Allen-Dynes modified McMillan formula (Eq. 1) or using the direct solution of the Eliashberg equation (see Table 1), where for the Coulomb pseudopotential μ* we use the commonly accepted bracketing values of 0.10 and 0.15. Detailed information on the calculated superconducting properties of uranium hydrides is given in Table 1, spectral function $\alpha^2 F(\omega)$ for UH$_{7-9}$ is presented on Fig. S6. For UH$_7$ at 20 GPa, we find EPC coefficient of 0.83 resulting in $T_c$ in the range 44-54 K (57-66 K at 0 GPa, see Table 1). The structurally related U$_2$H$_{13}$ should display similar EPC coefficient and $T_c$ values. For UH$_8$ at 50 GPa we predict λ = 0.73 and $T_c$ in the range 23-33 K (46-55 K at 0 GPa). $P6_3/mmc$-UH$_9$ at 300 GPa has the lowest $T_c$ among the above considered hydrides – 20-31 K at 300.

**Conclusions**

In summary, using the USPEX evolutionary crystal structure prediction algorithm we found 14 new uranium hydrides, including hydrogen-rich high-temperature superconductors UH$_7$, UH$_8$, UH$_9$, U$_2$H$_{13}$, and U$_2$H$_{17}$. Their crystal structures are based on either f.c.c. or h.c.p. uranium sublattice and H$_8$ cubic clusters. Our high-pressure experiments have successfully produced UH$_5$ at the pressure of 5 GPa, UH$_7$ at 31 GPa, and UH$_{8+\delta}$ at 45 GPa, corroborating our predictions and confirming their reliability. New uranium hydrides have been identified by close match of theoretically calculated and experimental XRD patterns and equations of state. We predict UH$_{7-9}$ to be superconductors with maximum $T_c$ for UH$_7$, 54 K at 20 GPa. Superconducting uranium hydrides appear at unusually low pressures. Given dynamical stability of UH$_{7-8}$ at zero pressure, there is a possibility for them to exist as metastable phases at ambient pressure where their $T_c$ values will reach 57-66 K. Furthermore, given the presence of a pseudogap at the Fermi level for all UH$_{7-9}$ compounds, we expect doping to be effective in raising $T_c$. This and other works (Ac-H (*19*), Th-H (*20*)) bring the possibility of new high-temperature superconductors based on actinides. Present work clearly shows the predictive power of modern methods of crystal structure prediction, capable of finding unusual materials with exotic chemistry.

**Materials and Methods:**

Theoretical calculations. The USPEX evolutionary algorithm (*14–16*) is a powerful tool for predicting thermodynamically stable compounds of given elements at a given pressure. We performed variable-composition searches in the U-H system at 0, 5, 25, 50, 100, 200, 300, 400, and 500 GPa. The first generation (120 structures) was created using a random symmetric generator, while all subsequent generations contained 20% of random structures and 80% of structures created using heredity, softmutation and transmutation operators. Within USPEX runs, structure relaxations were performed at the GGA level (with the functional from Ref. (*21*)) of density functional theory (DFT) using the projector-augmented wave (PAW) method (*22*) as implemented in the VASP code (*23–25*). Plane wave kinetic energy cutoff was set to 600 eV and the Brillouin zone was sampled by the Γ-centered k-mesh with a resolution of $2\pi \times 0.05$ Å$^{-1}$.

In order to establish stability fields of the predicted phases, we recalculated their enthalpies with increased precision at various pressures with a smaller pressure increment (from 1 to 10 GPa), recalculating the thermodynamic convex hull (Maxwell construction) at each pressure. The phases that were located on the convex hull are the ones stable at given pressure. Stable structures of elemental H and U were taken from USPEX calculations and from (*26*) and (*27*), respectively.

The superconducting $T_c$ was calculated using QUANTUM ESPRESSO package (*28*). The phonon frequencies and EPC coefficients were computed using density-functional perturbation theory (*29*) employing the plane-wave pseudopotential method and Perdew-Burke-Ernzerhof exchange-correlation functional (*21*). Convergence tests showed that 60 Ry is a suitable kinetic energy cutoff for the plane wave basis set. Electronic band structures of $UH_7$ and $UH_8$ were calculated using both VASP and QE and demonstrated good consistency. Comparison of the phonon densities of states calculated using the finite displacement method (VASP and PHONOPY (*30*)) and density-functional perturbation theory (QE) showed perfect agreement between these methods.

Critical temperature was calculated from the Eliashberg equation (*31*) which is based on the Fröhlich Hamiltonian $\hat{H} = \hat{H}_e + \hat{H}_{ph} + \sum_{k,q,j} g_{k+q,k}^{q,j} \hat{c}_{k+q}^+ \hat{c}_k (\hat{b}_{-q,j}^+ + \hat{b}_{q,j})$, where $c^+$, $b^+$ relate to creation operators of electrons and phonons, respectively. Matrix element of electron-phonon interaction $g_{k+q,k}^{q,j}$ calculated within the harmonic approximation in Quantum ESPRESSO can be defined as $g_{k+q,k}^{q,j} = \sqrt{\frac{\hbar}{2M\omega_{q,j}}} \int \psi_k^*(r) \cdot \left\{ \frac{dV_{scf}}{d\vec{u}_q} \cdot \frac{\vec{u}_q}{|\vec{u}_q|} \right\} \cdot \psi_{k+q}(r) d^3r$, where $u_q$ is the displacement of an atom with mass *M* in the phonon mode *q,j*. Within the framework of Gor'kov and Migdal approach (*32, 33*) the correction to the electron Green's function $\Sigma(\vec{k},\omega) = G_0^{-1}(\vec{k},\omega) - G^{-1}(\vec{k},\omega)$ caused by interaction can be calculated by taking into account only the first terms of the expansion of electron-phonon interaction in series of ($\omega_{log}/E_F$). As a result, it will lead to integral Eliashberg equations (*31*). These equations can be solved by iterative self-consistent method for the real part of the order parameter Δ(T, ω) (superconducting gap) and the mass renormalization function Z(T, ω) (*34*) (for more details see Supplementary Materials).

In our calculations of the electron-phonon coupling (EPC) parameter λ, the first Brillouin zone was sampled using a 2×2×2 q-points mesh and a denser 24×24×24 k-points mesh (with Gaussian smearing and σ = 0.03 Ry which approximates the zero-width limits in the calculation of λ). The superconducting transition temperature $T_c$ was estimated using the Allen-Dynes modified McMillan equation (*35*).

$$T_C = \frac{\omega_{log}}{1.2} exp\left(\frac{-1.04(1 + \lambda)}{\lambda - \mu^*(1 + 0.62\lambda)}\right), \qquad (1)$$

where $\omega_{log}$ is the logarithmic average frequency and $\mu^*$ is the Coulomb pseudopotential, for which we used widely accepted lower and upper bound values of 0.10 and 0.15. The EPC constant λ and $\omega_{log}$ were calculated as

$$\lambda = 2\int_0^\infty \frac{\alpha^2 F(\omega)}{\omega} d\omega \qquad (2)$$

and

$$\omega_{log} = exp\left(\frac{2}{\lambda}\int \frac{d\omega}{\omega} \alpha^2 F(\omega) ln(\omega)\right). \qquad (3)$$

Experiments. We performed experiments in laser-heated diamond anvil cells (DAC) with 200 to 300 μm central culets. Small pieces of uranium metal (with naturally oxidized surfaces) were thinned down to 5-8 μm and mechanically or laser-cut to 40-60 linear dimensions. These were positioned in a hole in a rhenium gasket and filled with $H_2$ gas at ≈ 150 MPa along with small Au chips for pressure measurements. In a control experiment the sample cavity was filled by Ar gas. The loaded material was successively laser-heated to 1700 K at various pressures during sample loading using microsecond long pulses of a 1064 nm Yb-doped YAG fiber laser (*36*). We used this pulsed laser heating mode in order to avoid premature diamond breakage which is common with DAC loaded with hydrogen. Synchrotron X-ray diffraction (XRD) measurements (X-ray wavelength 0.3344 Å) at GSECARS, APS, ANL (*37*) and an X-ray beam spot as small as 3×4 μm were used to probe the physical and chemical state of the sample.

**References and Notes:**
1. D. Duan *et al.*, Pressure-induced metallization of dense (H2S)2H2 with high-$T_c$ superconductivity. *Sci. Rep.* **4**, 6968 (2015).
2. A. P. Drozdov, M. I. Eremets, I. A. Troyan, V. Ksenofontov, S. I. Shylin, Conventional superconductivity at 203 kelvin at high pressures in the sulfur hydride system. *Nature*. **525**, 73–76 (2015).
3. F. Le Guyadec *et al.*, Pyrophoric behaviour of uranium hydride and uranium powders. *J. Nucl. Mater.* **396**, 294–302 (2010).
4. R. Orr *et al.*, Kinetics of the reaction between water and uranium hydride prepared under conditions relevant to uranium storage. *J. Alloys Compd.* **695**, 3727–3735 (2017).
5. S. Imoto, T. Tanabe, K. Utsunomiya, Separation of hydrogen isotopes with uranium hydride. *Int. J. Hydrogen Energy*. **7**, 597–601 (1982).
6. J. E. Burke, C. S. Smith, The Formation of Uranium Hydride. *J. Am. Chem. Soc.* **69**, 2500–2502 (1947).
7. R. N. R. Mulford, F. H. Ellinger, W. H. Zachari-Asen, A New Form of Uranium Hydride. *J. Am. Chem. Soc.* **76**, 297–298 (1954).
8. R. E. Rundle, The Structure of Uranium Hydride and Deuteride. *J. Am. Chem. Soc.* **69**, 1719–1723 (1947).
9. R. E. Rundle, The Hydrogen Positions in Uranium Hydride by Neutron Diffraction. *J. Am. Chem. Soc.* **73**, 4172–4174 (1951).
10. P. F. Souter, G. P. Kushto, L. Andrews, M. Neurock, Experimental and theoretical evidence for the formation of several uranium hydride molecules. *J. Am. Chem. Soc.* **119**, 1682–1687 (1997).
11. D. Aoki, J. Flouquet, Ferromagnetism and Superconductivity in Uranium Compounds. *J. Phys. Soc. Japan*. **81**, 011003 (2012).
12. D. Aoki *et al.*, Coexistence of superconductivity and ferromagnetism in URhGe. *Nature*. **413**, 613–616 (2001).
13. S. S. Saxena *et al.*, Superconductivity on the border of itinerant-electron ferromagnetism in UGe2. *Nature*. **406**, 587–592 (2000).
14. A. R. Oganov, C. W. Glass, Crystal structure prediction using *ab initio* evolutionary techniques: Principles and applications. *J. Chem. Phys.* **124**, 244704 (2006).
15. A. O. Lyakhov, A. R. Oganov, H. T. Stokes, Q. Zhu, New developments in evolutionary structure prediction algorithm USPEX. *Comput. Phys. Commun.* **184**, 1172–1182 (2013).
16. A. R. Oganov, A. O. Lyakhov, M. Valle, How Evolutionary Crystal Structure Prediction Works—and Why. *Acc. Chem. Res.* **44**, 227–237 (2011).
17. M. Idiri, T. Le Bihan, S. Heathman, J. Rebizant, Behavior of actinide dioxides under pressure: UO2 and ThO 2. *Phys. Rev. B - Condens. Matter Mater. Phys.* **70**, 014113 (2004).
18. T. Le Bihan *et al.*, Structural behavior of α-uranium with pressures to 100 GPa. *Phys. Rev. B*. **67**, 134102 (2003).
19. D. V. Semenok, A. G. Kvashnin, I. A. Kruglov, A. R. Oganov, Actinium Hydrides $AcH_{10}$, $AcH_{12}$, and $AcH_{16}$ as High-Temperature Conventional Superconductors. *J. Phys. Chem. Lett.*, 1920–1926 (2018).
20. A. G. Kvashnin, D. V. Semenok, I. A. Kruglov, A. R. Oganov, High-Temperature Superconductivity in Th-H System at Pressure Conditions (2017) (available at http://arxiv.org/abs/1711.00278).

21. J. P. Perdew, K. Burke, M. Ernzerhof, Generalized Gradient Approximation Made Simple. *Phys. Rev. Lett.* **77**, 3865–3868 (1996).
22. G. Kresse, D. Joubert, From ultrasoft pseudopotentials to the projector augmented-wave method. *Phys. Rev. B*. **59**, 1758–1775 (1999).
23. G. Kresse, J. Furthmüller, Efficient iterative schemes for *ab initio* total-energy calculations using a plane-wave basis set. *Phys. Rev. B*. **54**, 11169–11186 (1996).
24. G. Kresse, J. Hafner, *Ab initio* molecular dynamics for liquid metals. *Phys. Rev. B*. **47**, 558–561 (1993).
25. G. Kresse, J. Hafner, *Ab initio* molecular-dynamics simulation of the liquid-metal–amorphous-semiconductor transition in germanium. *Phys. Rev. B*. **49**, 14251–14269 (1994).
26. C. J. Pickard, R. J. Needs, Structure of phase III of solid hydrogen. *Nat. Phys.* **3**, 473–476 (2007).
27. S. Adak, H. Nakotte, P. F. de Châtel, B. Kiefer, Uranium at high pressure from first principles. *Phys. B Condens. Matter*. **406**, 3342–3347 (2011).
28. P. Giannozzi *et al.*, QUANTUM ESPRESSO: a modular and open-source software project for quantum simulations of materials. *J. Phys. Condens. Matter*. **21**, 395502 (2009).
29. S. Baroni, S. de Gironcoli, A. Dal Corso, P. Giannozzi, Phonons and related crystal properties from density-functional perturbation theory. *Rev. Mod. Phys.* **73**, 515–562 (2001).
30. A. Togo, I. Tanaka, First principles phonon calculations in materials science. *Scr. Mater.* **108**, 1–5 (2015).
31. G. M. Eliashberg, Interactions between Electrons and Lattice Vibrations in a Superconductor. *JETP*. **11**, 696–702 (1959).
32. L. P. Gor'kov, On energy spectrum of superconductors. *JETP*. **34**, 735 (1958).
33. A. B. Migdal, Interaction between electrons and lattice vibrations in a normal metal. *JETP*. **34**, 996–1001 (1958).
34. E. G. Maksimov, D. Y. Savrasov, S. Y. Savrasov, The electron-phonon interaction and the physical properties of metals. *Physics-Uspekhi*. **40**, 337 (1997).
35. P. B. Allen, R. C. Dynes, Transition temperature of strong-coupled superconductors reanalyzed. *Phys. Rev. B*. **12**, 905–922 (1975).
36. A. F. Goncharov *et al.*, X-ray diffraction in the pulsed laser heated diamond anvil cell. *Rev. Sci. Instrum.* **81**, 113902 (2010).
37. V. B. Prakapenka *et al.*, Advanced flat top laser heating system for high pressure research at GSECARS: application to the melting behavior of germanium. *High Press. Res.* **28**, 225–235 (2008).

**Acknowledgments**

This work was supported by Russian Science Foundation (grant 16-13-10459). Portions of this work were performed at GeoSoilEnviroCARS (The University of Chicago, Sector 13), Advanced Photon Source (APS), Argonne National Laboratory. GeoSoilEnviroCARS is supported by the National Science Foundation - Earth Sciences (EAR - 1634415) and Department of Energy-GeoSciences (DE-FG02-94ER14466). This research used resources of the Advanced Photon Source, a U.S. Department of Energy (DOE) Office of Science User Facility operated for the DOE Office of Science by Argonne National Laboratory under Contract No. DE-AC02-06CH11357. Financial support from the National Natural Science Foundation of China (Nos. 11674330, 11504382, and 11604342) is gratefully acknowledged. A.F.G. was supported by the Chinese Academy of Sciences visiting professorship for senior international scientists (Grant No. 2011T2J20) and Recruitment Program of Foreign Experts.

**Supplementary Materials**

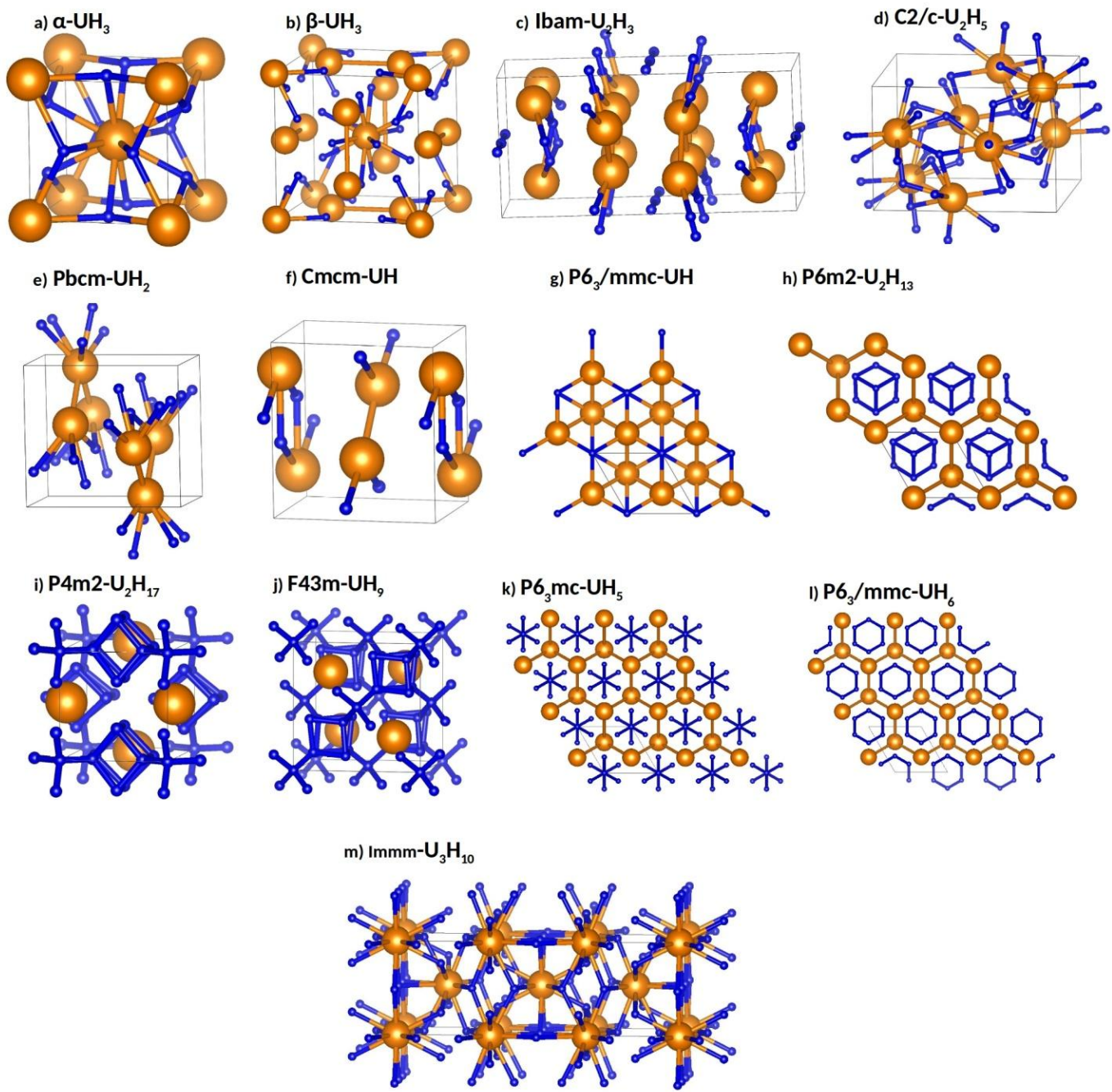

Figure S1: Crystal structures of the predicted U-H phases. U atoms are shown by large orange balls and hydrogens by small blue balls.

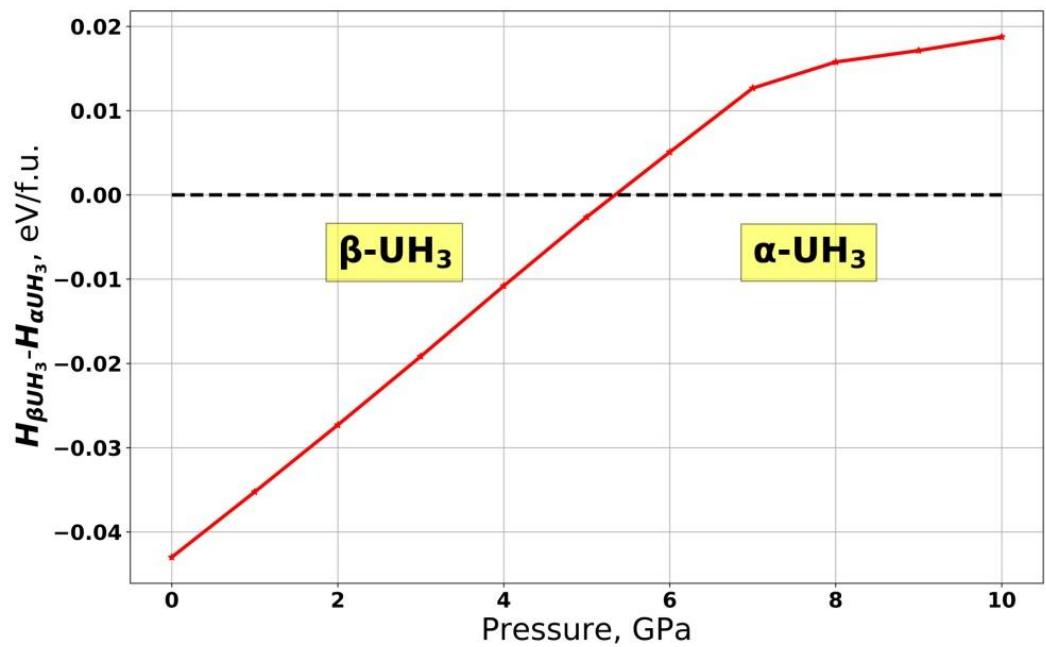

Figure S2: Enthalpy difference between β-UH$_3$ and α-UH$_3$.

Table S1: Crystal structures of predicted uranium hydrides

| Phase | Space group | Lattice parameters, Å | Atom | x | y | z |
|---|---|---|---|---|---|---|
| UH (30 GPa) | $Cmcm$ | a = 4.637<br>b = 4.945<br>c = 3.548 | U1<br>H1 | 0.0<br>0.0 | 0.1816<br>-0.3995 | 0.25<br>0.25 |
| UH (65 GPa) | $P6_3/mmc$ | a = 3.478<br>c = 3.361 | U1<br>H1 | 0.3333<br>0.0 | 0.6667<br>0.0 | 0.75<br>0.0 |
| $U_2H_3$ (30 GPa) | $Ibam$ | a = 10.175<br>b = 4.881<br>c = 3.537 | U1<br>H1<br>H2 | -0.3758<br>-0.1701<br>0.5 | -0.3121<br>-0.3975<br>0.0 | 0.0<br>0.0<br>0.25 |
| $UH_2$ (30 GPa) | $Pbcm$ | a = 5.398<br>b = 4.867<br>c = 3.589 | U1<br>H1<br>H2 | 0.2617<br>0.3497<br>0.0513 | -0.0616<br>0.3554<br>0.25 | 0.25<br>0.25<br>0.0 |
| $U_2H_5$ (30 GPa) | $C2/c$ | a = 5.164<br>b = 5.232<br>c = 7.480<br>β = 96.18° | U1<br>H1<br>H2<br>H3 | -0.2174<br>-0.3388<br>0.1036<br>0.0 | 0.0501<br>0.3237<br>0.4299<br>0.2247 | -0.0966<br>0.3586<br>-0.3504<br>0.25 |
| α-$UH_3$ (0 GPa) | $Pm\bar{3}n$ | a = 4.118 | U1<br>H1 | 0.0<br>0.25 | 0.0<br>0.5 | 0.0<br>0.0 |
| β-$UH_3$ (0 GPa) | $Pm\bar{3}n$ | a = 6.576 | U1<br>U2<br>H1 | 0.0<br>0.25<br>0.0 | 0.0<br>0.0<br>0.1552 | 0.0<br>0.5<br>-0.3047 |
| $U_3H_{10}$ (20 GPa) | $Immm$ | a = 3.836<br>b = 12.089<br>c = 3.997 | U1<br>U2<br>H1<br>H2<br>H3<br>H4 | 0.0<br>0.0<br>0.0<br>-0.2467<br>0.0<br>0.0 | 0.0<br>-0.3310<br>0.1565<br>0.5<br>-0.4160<br>-0.2663 | 0.0<br>0.0<br>0.2375<br>0.0<br>0.5<br>0.5 |
| $UH_5$ (20 GPa) | $P6_3mc$ | a = 3.695<br>c = 5.816 | U1<br>H1<br>H2<br>H3 | 0.3333<br>0.0<br>-0.1720<br>0.3333 | 0.6667<br>0.0<br>0.1720<br>0.6667 | -0.4007<br>0.1370<br>0.4104<br>0.2240 |
| $UH_6$ (20 GPa) | $P6_3/mmc$ | a = 3.944<br>c = 5.385 | U1<br>H1 | 0.3333<br>-0.1772 | 0.6667<br>-0.3543 | 0.25<br>0.0973 |

| Compound | Space group | Lattice parameters (Å) | Atom | x | y | z |
|---|---|---|---|---|---|---|
| U$_2$H$_{13}$ (20 GPa) | $P\bar{6}m2$ | a = 3.908<br>c = 5.521 | U1<br>U2<br>H1<br>H2<br>H3 | 0.6667<br>0.0<br>0.1506<br>-0.4907<br>0.3333 | 0.3333<br>0.0<br>-0.1506<br>0.4907<br>0.6667 | 0.0<br>0.5<br>-0.1691<br>0.3455<br>0.0 |
| UH$_7$ (20 GPa) | $P6_3/mmc$ | a = 3.894<br>c = 5.645 | U1<br>H1<br>H2 | 0.3333<br>0.0<br>-0.1822 | 0.6667<br>0.0<br>-0.3644 | 0.25<br>0.25<br>0.0789 |
| UH$_8$ (50 GPa) | $Fm\bar{3}m$ | a = 5.158 | U1<br>H1 | 0.0<br>-0.3661 | 0.0<br>-0.3661 | 0.0<br>-0.3661 |
| U$_2$H$_{17}$ (300 GPa) | $P\bar{4}m2$ | a = 3.235<br>c = 4.578 | U1<br>H1<br>H2<br>H3<br>H4<br>H5 | 0.0<br>0.2829<br>-0.2687<br>-0.2301<br>-0.2753<br>0.0 | 0.5<br>0.5<br>0.0<br>0.0<br>0.5<br>0.0 | -0.2498<br>0.1135<br>-0.1389<br>-0.3902<br>0.3897<br>0.0 |
| UH$_9$ (300 GPa) | $P6_3/mmc$ | a = 3.190<br>c = 5.553 | U1<br>H1<br>H2<br>H3 | 0.3333<br>0.1845<br>0.0<br>0.3333 | 0.6667<br>0.3690<br>0.0<br>0.6667 | 0.25<br>-0.0620<br>0.3484<br>0.75 |
| UH$_9$ (300 GPa) | $F\bar{4}3m$ | a = 4.609 | U1<br>H1<br>H2<br>H3 | 0.25<br>-0.3915<br>-0.1321<br>0.0 | 0.25<br>-0.3915<br>-0.1321<br>0.0 | 0.25<br>-0.3915<br>-0.1321<br>0.0 |

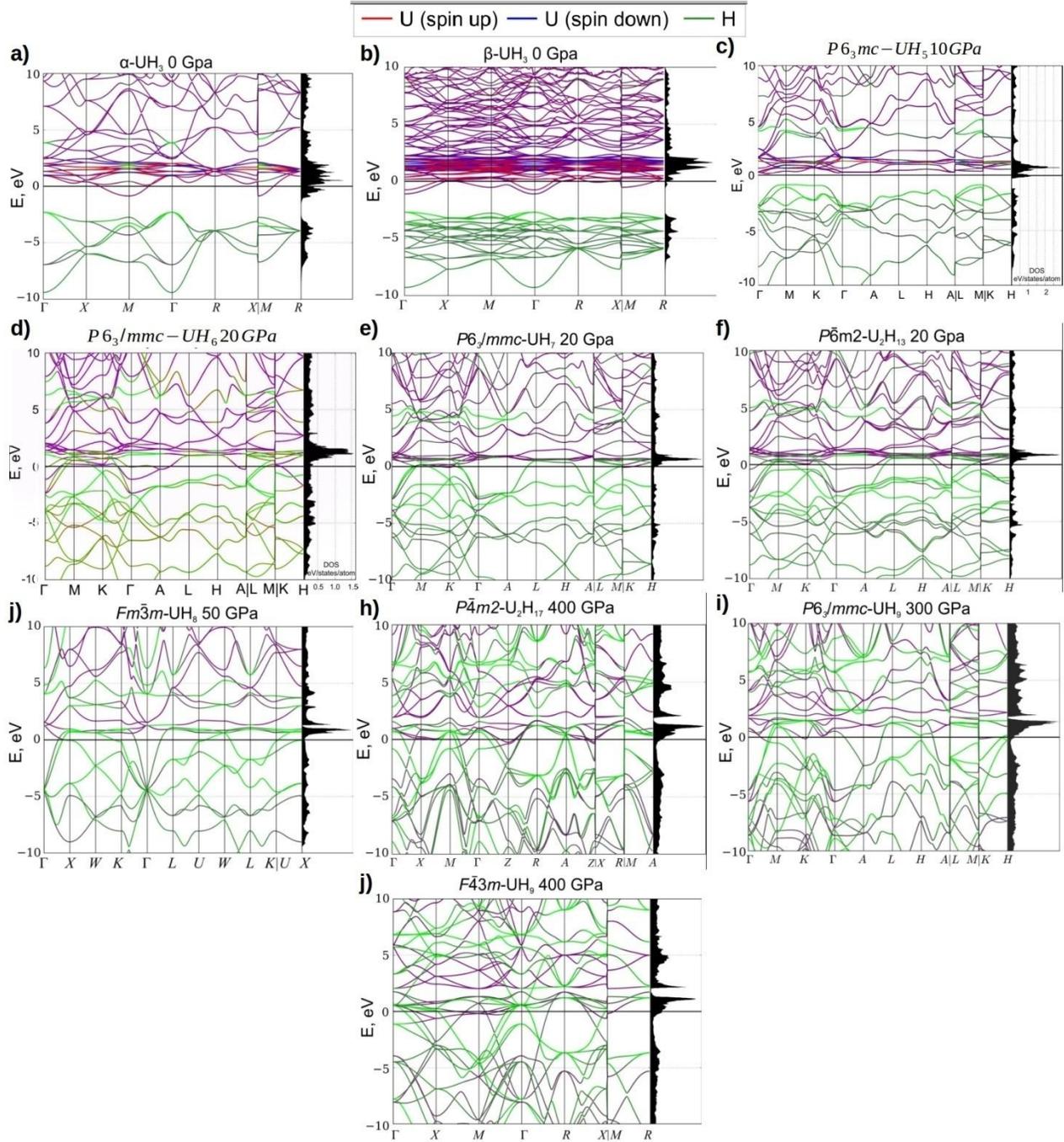

Figure S3: Electronic band structures of predicted uranium hydrides. Spin-up and spin-down contributions of uranium orbitals are shown in red and blue, respectively, and the contribution of hydrogen in green.

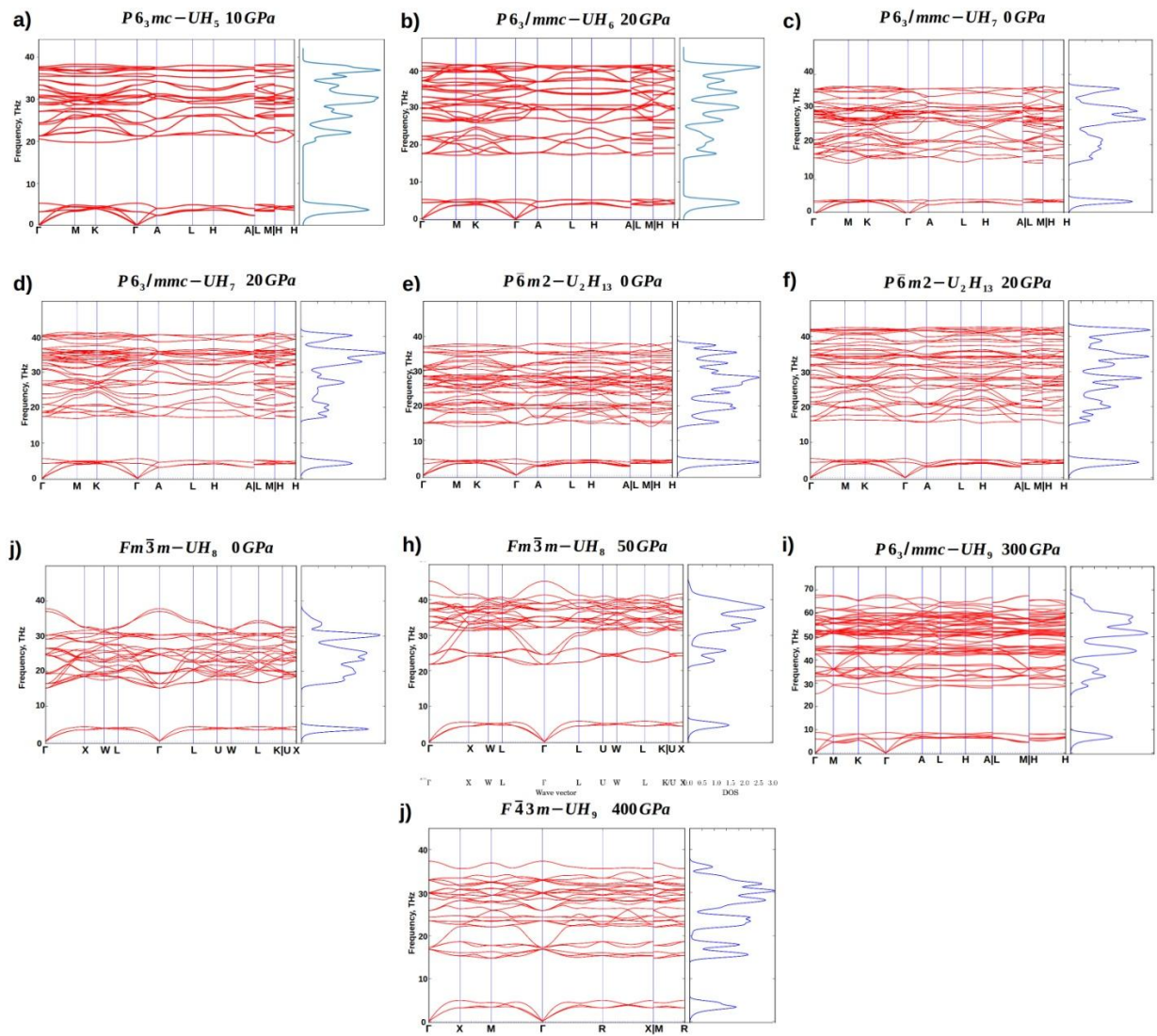

Figure S4: Phonon band structures and densities of states of predicted uranium hydrides.

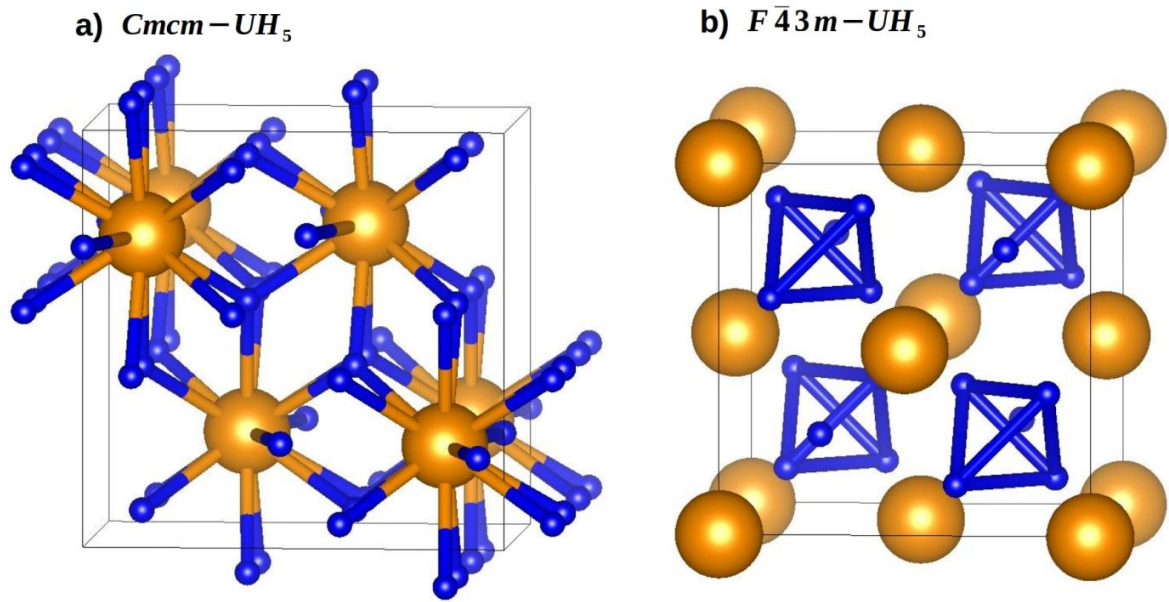

Figure S5: Crystal structures of the experimentally found UH$_5$ structures.

Table S2. Predicted superconducting properties of uranium hydrides. Two $T_c$ values are given for $\mu^*$ equal to 0.1 and 0.15, respectively.

| Phase | Space group | P, GPa | $\omega_{log}$, K | $\lambda$ | $T_c$, K |
|---|---|---|---|---|---|
| UH$_7$ | $P6_3/mmc$ | 20 | 873.8 | 0.83 | 47.6 / 32.7 |
| | | 0 | 764.9 | 0.95 | 57.5 / 49.3 |
| UH$_8$ | $Fm\bar{3}m$ | 50 | 873.7 | 0.73 | 27.5 / 22.2 |
| | | 0 | 450.3 | 1.13 | 37.6 / 30.2 |
| UH$_9$ | $P6_3/mmc$ | 300 | 933.4 | 0.67 | 35.8 / 20.8 |

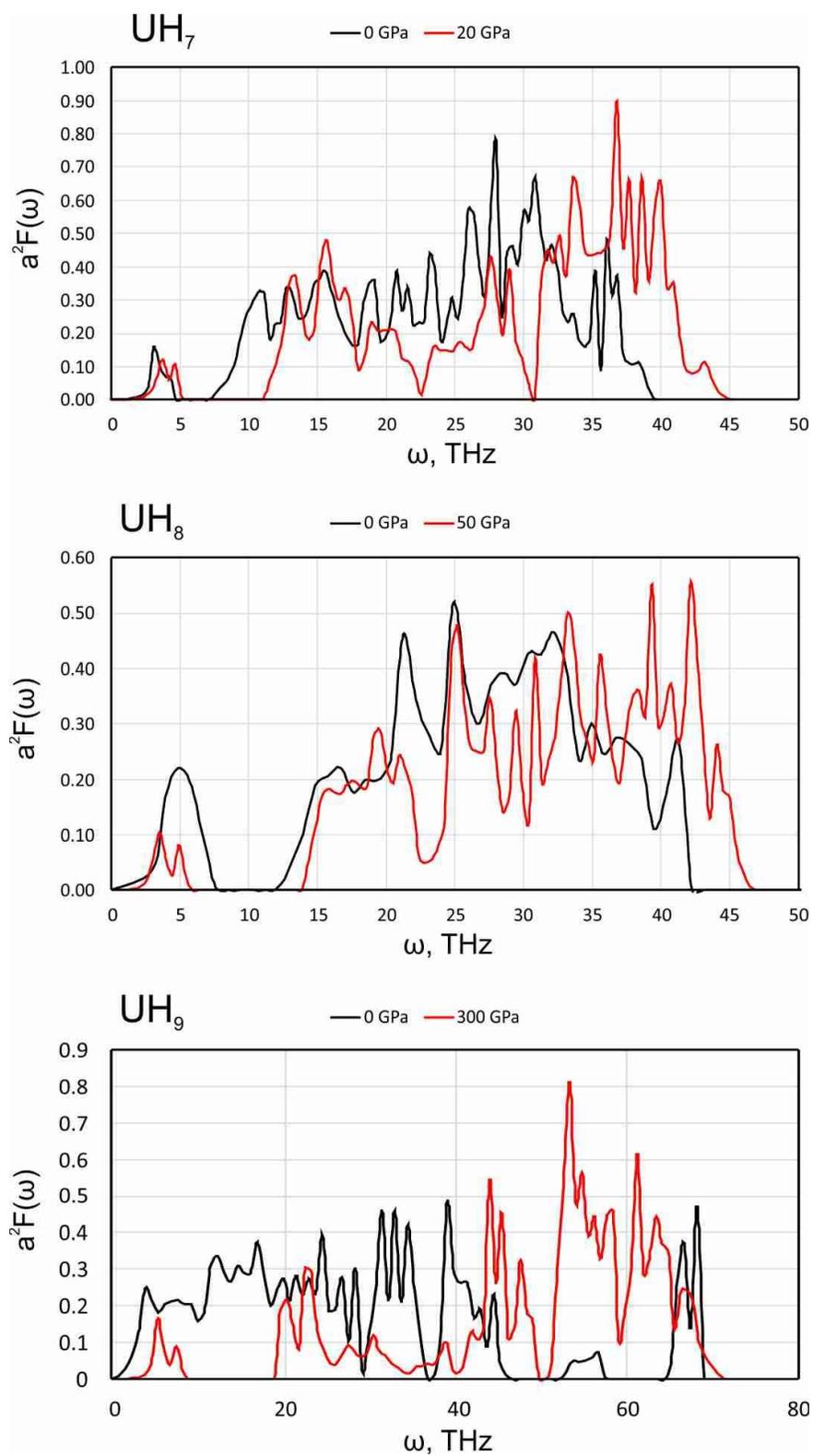

Figure S6: Spectral function $\alpha^2 F(\omega)$ for different uranium hydrides.

## Solving of the Eliashberg equation

At the first stage of self-consistent procedure of the numerical solution of the Eliashberg equation (*1*) we calculated Fredholm kernels of electron-phonon interaction $I_1(\omega,\omega')$, $I_{-1}(\omega,\omega')$, $I_2(\omega,\omega')$. Integration was divided into intervals using the following symmetry properties $\Delta(-\omega) = \Delta^*(\omega)$, $Z(-\omega) = Z^*(\omega)$, $\alpha^2(-\omega)F(-\omega) = \alpha^2(\omega)F(\omega)$(*2*), but avoiding points T = 0 K, $\omega$ = 0 Ry due to possible divergences of 1/T, 1/$\omega$, $\omega$/T during numerical integration (*2–4*):

$$I_1(\omega,\omega') = \int_0^\infty \alpha^2(\Omega)F(\Omega) \cdot \left\{ \frac{\left(\exp\left(-\frac{\hbar\omega'}{kT}\right)+1\right)^{-1} + \left(\exp\left(\frac{\hbar\Omega}{kT}\right)-1\right)^{-1}}{\hbar\omega'+\hbar\Omega-\hbar\omega} + \frac{\left(\exp\left(\frac{\hbar\omega'}{kT}\right)+1\right)^{-1} + \left(\exp\left(\frac{\hbar\Omega}{kT}\right)-1\right)^{-1}}{\hbar\omega'-\hbar\Omega-\hbar\omega} \right\} \cdot \hbar d\Omega$$

$$I_{-1}(\omega,\omega') = \int_0^\infty \alpha^2(\Omega)F(\Omega) \cdot \left\{ \frac{\left(\exp\left(\frac{\hbar\omega'}{kT}\right)+1\right)^{-1} + \left(\exp\left(\frac{\hbar\Omega}{kT}\right)-1\right)^{-1}}{-\hbar\omega'+\hbar\Omega-\hbar\omega} - \frac{\left(\exp\left(-\frac{\hbar\omega'}{kT}\right)+1\right)^{-1} + \left(\exp\left(\frac{\hbar\Omega}{kT}\right)-1\right)^{-1}}{\hbar\omega'+\hbar\Omega+\hbar\omega} \right\} \cdot \hbar d\Omega$$

$$\frac{I_2(\omega,\omega')}{\hbar\omega} = \int_0^\infty 2\alpha^2(\Omega)F(\Omega) \cdot \left\{ \frac{\left(\exp\left(\frac{\hbar\omega'}{kT}\right)+1\right)^{-1} + \left(\exp\left(\frac{\hbar\Omega}{kT}\right)-1\right)^{-1}}{(\hbar\omega)^2 - (\hbar\Omega-\hbar\omega')^2} - \frac{\left(\exp\left(-\frac{\hbar\omega'}{kT}\right)+1\right)^{-1} + \left(\exp\left(\frac{\hbar\Omega}{kT}\right)-1\right)^{-1}}{(\hbar\Omega+\hbar\omega')^2 - (\hbar\omega)^2} + \right\} \cdot \hbar d\Omega$$

Then, we calculated the Coulomb contribution to the order parameter using common expression for electronic density of states in BCS theory(*2*):

$$I_3 = -\mu \cdot \int_0^\infty \mathrm{Re}\left(\frac{\Delta(\omega')}{\sqrt{\hbar^2\omega'^2 - \Delta^2(\omega')}}\right) \cdot \frac{\exp\left(\frac{\hbar\omega'}{kT}\right) - 1}{\exp\left(\frac{\hbar\omega'}{kT}\right) + 1} \cdot \hbar d\omega'.$$

Calculation of average Coulomb electron-electron repulsion was performed using the upper bound of empirical value $\mu^* = 0.1$ (0.15) with $\omega_c$ = 75 THz = 0.31 eV (upper bound of phonon spectrum) and $E_e$ = 85 eV (characteristic width of electron spectrum in AcH$_{10}$) by a common way gives $\mu$ = 0.23 (0.95).

$$\mu = \frac{\mu^*}{1 - \mu^* \ln\frac{E_e}{\hbar\omega_c}}$$

Then, a function of electron mass renormalization $Z(\omega)$ was calculated as

$$Z(\omega) = 1 - \int_0^\infty \mathrm{Re}\left(\frac{\hbar\omega'}{\sqrt{\hbar^2\omega'^2 - \Delta^2(\omega')}}\right) \cdot \frac{I_2(\omega,\omega')}{\hbar\omega} \cdot \hbar d\omega',$$

After that the next approximation for the order parameter (or the value of superconducting gap, $\Delta(\omega)$)

$$\Delta(\omega) = \frac{I_3 + \int_0^\infty \mathrm{Re}\left(\frac{\Delta(\omega')}{\sqrt{\hbar^2\omega'^2 - \Delta^2(\omega')}}\right) \cdot I_1(\omega,\omega')\,\hbar d\omega' + \int_0^\infty \mathrm{Re}\left(\frac{\Delta(\omega')}{\sqrt{\hbar^2\omega'^2 - \Delta^2(\omega')}}\right) \cdot I_{-1}(\omega,\omega')\,\hbar d\omega'}{Z(\omega)}$$

and the new $I_1(\omega,\omega')$, $I_{-1}(\omega,\omega')$, $I_2(\omega,\omega')$ integrals are calculated.

Then, the next iteration for $\Delta(\omega)$ is calculated and the new cycle starts again. Value of $\Delta(\omega)$ obtained after 10-20 iterations was used for construction of $\Delta(T, \omega)|_{\omega \to 0}$ = $\Delta(T)$ function and search for Tc ($\Delta$(Tc) →0) as well as characteristic ratio 2$\Delta$(0)/Tc and its deviation from weak coupling limit (3.52).

**References**
1.	G. M. Eliashberg, Interactions between Electrons and Lattice Vibrations in a Superconductor. *JETP*. **11**, 696–702 (1959).
2.	A. E. Karakozov, E. . Maksimov, A. A. Mikhailovskii, Superconductivity in systems with strong electron-phonon interaction. *JETP*. **75**, 70–76 (1992).
3.	N. A. Kudryashov, A. A. Kutukov, E. A. Mazur, Critical temperature of metallic hydrogen at a pressure of 500 GPa. *Jetp Lett.* **104**, 460–465 (2016).
4.	N. A. Kudryashov, A. A. Kutukov, E. A. Mazur, Critical temperature of metallic sulfur hydride at 225 GPa. *ZhETF*. **151**, 165 (2017).